\documentclass[aps,prl,twocolumn,floatfix,amsmath,superscriptaddress,nofootinbib]{revtex4}

\usepackage{bm}
\usepackage{graphicx}
\usepackage{latexsym}
\usepackage{amssymb}
\usepackage{amsmath}
\usepackage{color}
\usepackage{hyperref}
\usepackage{braket}
\usepackage{booktabs}
\usepackage{multirow}
\usepackage{soul}

\usepackage{siunitx}

\newcommand{\s}[2]{\sigma_{#2}^{#1}}

\newcommand{\darr}{{\downarrow}}
\newcommand{\uarr}{{\uparrow}}

\newcommand{\be}{\begin{equation}} 
\newcommand{\ee}{\end{equation}}
\newcommand{\ba}{\begin{array}}
	\newcommand{\ea}{\end{array}}
\newcommand{\bqa}{\begin{eqnarray}}
\newcommand{\eqa}{\end{eqnarray}}

\begin{document}

\title{Bayesian optimal control of GHZ states in Rydberg lattices}

\author{Rick Mukherjee}
\affiliation{Department of Physics, Imperial College London, SW7 2AZ, London, UK}
\author{Harry Xie}
\affiliation{Department of Physics, Imperial College London, SW7 2AZ, London, UK}
\author{Florian Mintert}
\affiliation{Department of Physics, Imperial College London, SW7 2AZ, London, UK}

\date{\today}
\begin{abstract}
The ability to prepare non-classical states in a robust manner is essential for quantum sensors beyond the standard quantum limit. We demonstrate that Bayesian optimal control is capable of finding control pulses that drive trapped Rydberg atoms into highly entangled GHZ states. The control sequences have a physically intuitive functionality based on the quasi-integrability of the Ising dynamics. They can be constructed in laboratory experiments resulting in preparation times that scale very favourably with the system size.
\end{abstract}

\maketitle

Among the variety of highly entangled many-body states, GHZ states~\cite{GHZorig}  are particularly useful with potential applications in quantum information~\cite{Shor, Steane, Knill}, quantum communication~\cite{Zhao}, cryptography~\cite{Hillery} and high precision interferometry~\cite{Leibfried_interf}. Experimental realizations include trapped ions~\cite{Monz}, photons~\cite{Ourjoumtsev83, Wang}, superconducting qubits~\cite{Song574, wei2019verifying} and nuclear spins~\cite{Neumann1326}. Most applications of GHZ states rely on their ability to perform as good sensors, but this sensitivity also makes them more vulnerable to external noise which ultimately destroys the many-body superposition state. Since both sensitivity and vulnerability grows with increasing system size \cite{Carvalho}, it is essential to prepare these states rapidly.

Rydberg atoms provide an excellent platform to create entangled states~\cite{Mukherjee_2011,Cui_2017, Khazali}. Bell states can be prepared with very high fidelity~\cite{Levine,madjarov2020highfidelity}, and a recently prepared GHZ state of 20 Rydberg atoms~\cite{Omran570} exceeds all previously realized GHZ states in number of constituents. Such state preparations are facilitated by the strong  Rydberg interactions~\cite{Tong, Robert} that are controllable~\cite{Robicheaux, Honer} in terms of the frequency and amplitude of the driving laser field. The interaction geometry can be controlled by trapping the atoms in optical lattices, which are routinely realized in one-~\cite{Schauss1455}, two-~\cite{Graham,Nogrette, wang2019preparation} and three-dimensional~\cite{barredo2019threedimensional} geometries.

As the number of involved Rydberg atoms increases, it is getting increasingly challenging to find suitable control fields that drive a system of interacting Rydberg atoms towards a desired state~\cite{Omran570}.  In particular the abundance of (close-to)-degenerate states make this control problem extremely demanding. Regularly utilized control algorithms may thus require additional control fields that lift degeneracies and show sufficiently good convergence for selected states only~\cite{Omran570}.

We show here that statistical machine learning techniques~\cite{frazier2018tutorial, sauvage2019} can increase the efficiency and fidelity of such a state preparation substantially. The main difference between statistical machine learning and control techniques that are more commonly used in quantum physics lie in the way in which the dependence of system dynamics on a control pulse are predicted. A probabilistic modeling based on Bayesian inference, allows statistical machine learning to converge very quickly \cite{Mukherjee_2020}, and it reduces the risk of getting trapped in local extrema of the control landscape, {\it i.e.} solutions that are only seemingly optimal. 

Considering different interaction geometries and target states, we will demonstrate that the difficulty of the control problem of state-preparation depends sensitively on the spectral properties of the underlying system, and that statistical machine learning can find very good solutions even in the presence of rather adverse spectra. On solving the control task at hand, we uncover a functional understanding of the control sequences which is to exploit the quasi-integrability of the Ising dynamics in order efficiently create the GHZ states. This is in contrast to the frequently encountered situation of highly complicated control pulses whose functionality eludes our intuition.

Finding temporal shapes of driving fields for large systems through numerical simulations is prohibitively difficult due to the numerical effort in simulating the dynamics, but optimizations based on experimental observations \cite{rabitz,Lu:2017aa,Dive2018} are a practical route towards efficient control of large systems. The optimizations for numerically accessible system sizes together with a figure of merit that can be estimated efficiently in an experiment, establish a promising methodology for state-of-the-art experiments~\cite{Nogrette,Schauss1455,barredo2019threedimensional}.

A general GHZ state is of the form
\be
\frac{1}{\sqrt{2}}\left(\ket{\alpha}+e^{i\phi}\ket{\beta}\right)\ ,
\label{eq:GHZ}
\ee
where $\ket{\alpha}$ and $\ket{\beta}$ are $N$-partite product states, such that each factor in $\ket{\alpha}$ is orthogonal to the corresponding factor in $\ket{\beta}$. In the following, we will consider the state $\ket{\Phi_N}$ with $\ket{\alpha}=\ket{\darr_1\darr_2\hdots \darr_N}$ and $\ket{\beta}=\ket{\uarr_1\uarr_2\hdots \uarr_N}$, and the state $\ket{\Psi_N}$ with $\ket{\alpha}=\ket{\darr_1\uarr_2\darr_3\hdots}$ and $\ket{\beta}=\ket{\uarr_1\darr_2\uarr_3\hdots}$.

The system of interacting Rydberg atoms driven by a laser field with an effective Rabi-frequency $\Omega(t)$ and detuning $\Delta(t)$ is described by the Hamiltonian
\begin{align}\label{Ising}
\hat{H}(t) &= \sum_i \frac{\hbar\omega_i(t)}{2}\s{z}{i}+\hbar\Omega(t) \sum_i \s{x}{i}  + \sum_{i<j} V_{ij} \s{z}{i} \s{z}{j}\ .
\end{align}
in the frame rotating with the driving field. The Pauli-matrices $\s{x}{i}$ and $\s{z}{i}$ of atom $i$ are defined in terms of the two levels $\ket{\darr_i}$ and $\ket{\uarr_i}$ that are driven close-to or in resonance by the driving field. The atomic resonance frequency $\omega_i(t)=2(\sum_j V_{ij}/\hbar-\Delta(t))$ contains a contribution of the interaction constants $V_{ij}= C_6(n)/|r_i-r_j|^{6}$ between Rydberg atoms $i$ and $j$. $C_6(n)$ is the van der Waals coefficient which depends on the principal quantum number $n$ of the excited atom. Since the interaction decays with the relative distance $|r_i-r_j|$ for each pair of atoms, the interaction landscape depends strongly on the trapping geometry.
 \begin{figure}[t!]
	\centering
	\includegraphics[width=1.0\columnwidth]{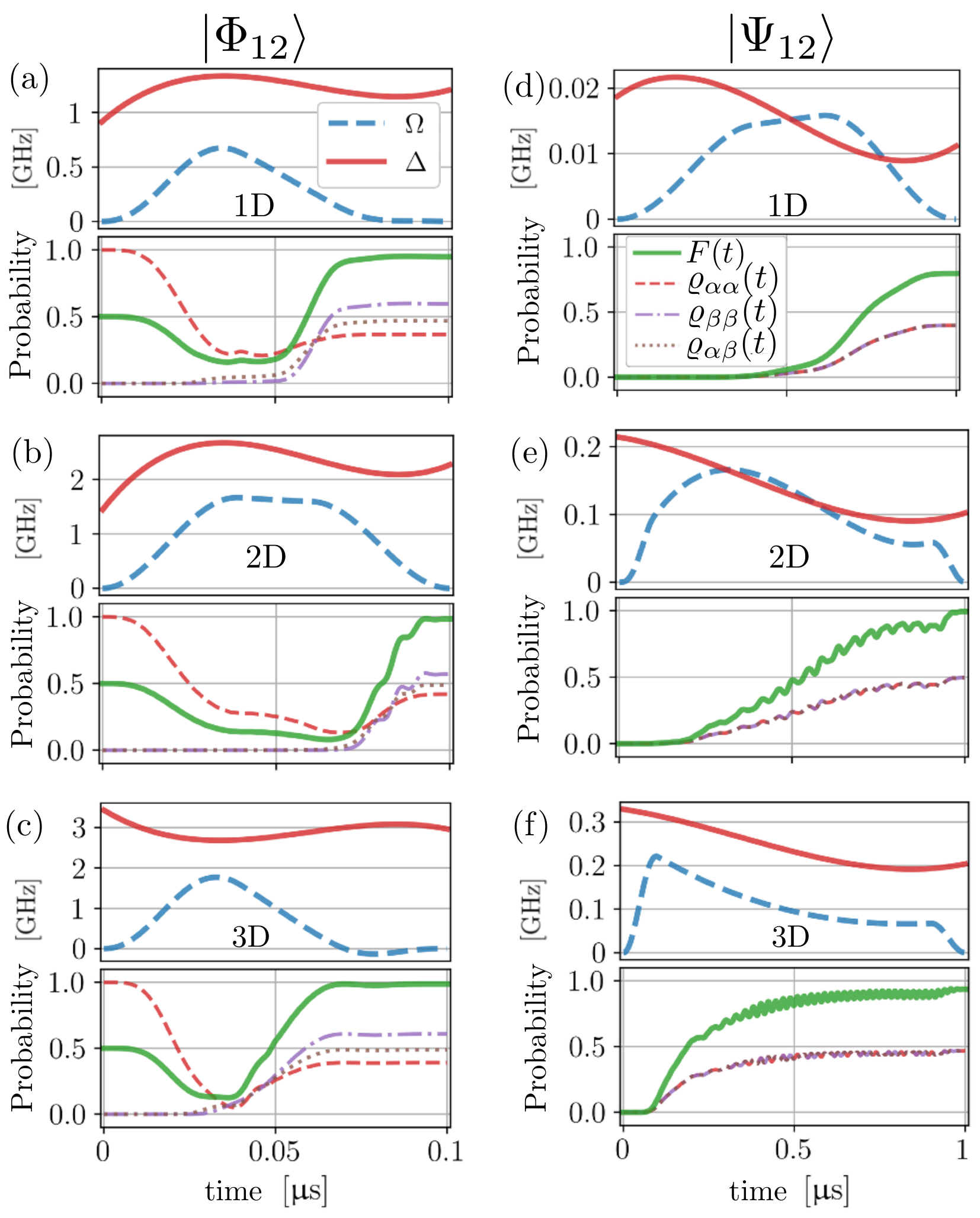}
	\caption{Optimized control fields $(\Omega(t), \Delta(t))$ (top panels) and dynamics of fidelity $F(t)$ and matrix elements (bottom panels) induced by the optimized control pulses.}
	\label{fiddyn}
\end{figure}

Since the relative phase $\phi$ in Eq.~\eqref{eq:GHZ}  can be adjusted in terms of single-particle dynamics without controlling any inter-Rydberg interaction, we will define the fidelity of a state $\varrho$ in terms of a maximization over the phase angle $\phi$, resulting in
\be\label{fid}
F(\varrho(t))=\frac{1}{2}\bigl(\varrho_{\alpha\alpha}(t)+\varrho_{\beta\beta}(t)\bigr)+\left|\varrho_{\alpha\beta}(t)\right| \ ,
\ee
with the probabilities $\varrho_{\alpha\alpha}(t)=\bra{\alpha}\varrho(t)\ket{\alpha}$ and $\varrho_{\beta\beta}(t)=\bra{\beta}\varrho(t)\ket{\beta}$ to find the system in state $\ket{\alpha}$ and $\ket{\beta}$, and the off-diagonal element $\varrho_{\alpha\beta}(t)=\bra{\alpha}\varrho(t)\ket{\beta}$ quantifying the coherence of the state.

The optimization problem at hand is the quest for a time-dependent driving frequency $\Omega(t)$ and detuning $\Delta(t)$ that results in the formation of the desired GHZ state\footnote{Optimization details provided in \cite{supp}.}.
Fig.~\ref{fiddyn} depicts optimal solutions for control pulses, population dynamics and fidelity (Eq.~\eqref{fid}) for the explicit example of twelve $^{87}$Rb atoms trapped in a two-dimensional geometry;
with ground state $\ket{\darr}=\ket{5S}$ and Rydberg state $\ket{\uarr} = \ket{50S}$ the Rydberg-Rydberg interactions are repulsive.
The sub-figures (a)-(c) and (d)-(e) correspond to the target states $\ket{\Phi_{12}}$ and $\ket{\Psi_{12}}$ respectively for different lattice dimensions.

The GHZ states $|\Phi\rangle$ and$|\Psi\rangle$ are created in durations of $0.1\mu s$ and $1 \mu s$ respectively for 12 qubits. Both durations are sufficiently fast so that decoherence is mostly negligible \cite{Saffman_rev}, and fidelities obtained in 2D and 3D lattices exceed the value of $0.99$. Given the importance of the state $|\Phi\rangle$ (that is yet to be realized in large Rydberg systems) for precision sensing, the present control techniques, promise crucial logistic benefit to real experiments~\cite{Omran570}.   

\begin{figure}[t!]
\centering
\includegraphics[width=1.0\columnwidth]{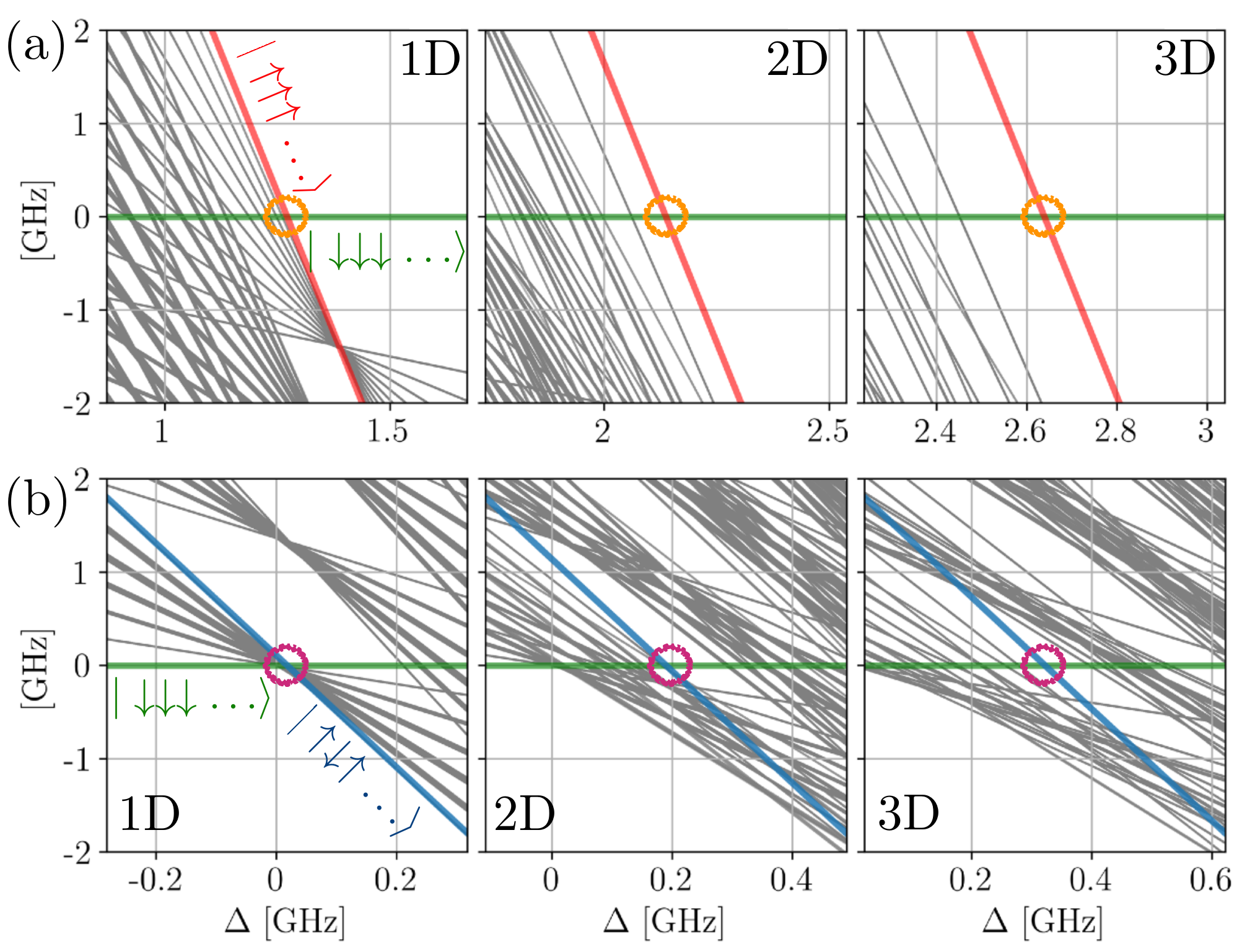}
\caption{(a-b) Level diagram of
12 Rb atoms with lattice spacing $l=1.5~\si{\micro}$m and Rydberg state $50S$ for different lattice dimensions in the zero-field limit ($\Omega(t)\to 0$). The eigenstates (green, red and blue bold lines) and crossings (orange and purple circles) of highest relevance for the state preparation are highlighted.}
	\label{manybodylevel}
\end{figure}

The functionality of the optimized dynamics can be understood in terms of the energy level structure of the system Hamiltonian. Fig.~\ref{manybodylevel} depicts the level diagram as a function of the detuning, in the limit of vanishing driving amplitude.
The levels that form the components for GHZ states $\ket{\Phi_{12}}$ and $\ket{\Psi_{12}}$ are emphasized (green, blue and red lines) in Fig.~\ref{manybodylevel}.
The inter-section of the GHZ state components with the initial state $|\darr\darr\darr\hdots\rangle$ are highlighted with (orange and purple) circles. 

Ideally, the preparation of a GHZ state could be realized in terms of a comparatively simple Landau-Zener transition if there were only the two levels $\ket{\alpha}$ and $\ket{\beta}$. In practice, however, many additional levels become populated, and a high-fidelity state can be obtained only if those populations vanish at the final point in time.  In higher lattice dimensions, these crossings occur for larger detunings where there are fewer undesired levels;
that is, fewer Landau-Zener transitions need to be controlled. 
Similarly, in all three interaction geometries, the crossing between the components of $\ket{\Phi_{N}}$ is at the far edge of the spectrum where there are substantially fewer undesired levels than in those parts of the spectrum in which the components of $\ket{\Psi_{N}}$ cross. This makes the preparation of $\ket{\Phi_{N}}$ a less demanding task than the preparation of $\ket{\Psi_{N}}$, so that it can be realised with faster sweeps, {\it i.e.} shorter times.

\begin{figure}[t!]
	\centering
	\includegraphics[width=1.0\columnwidth]{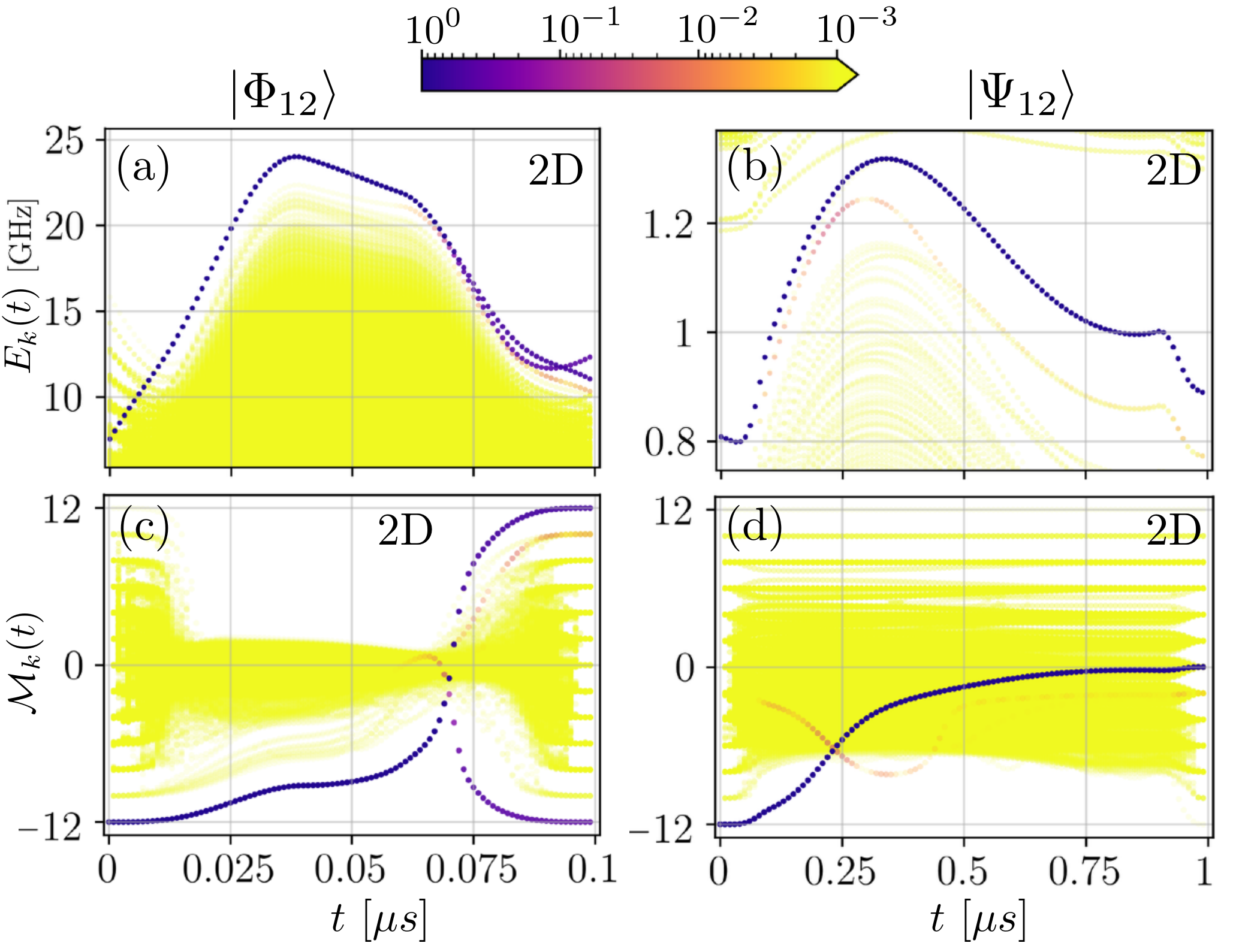}
	\caption{Energy (a and b) and magnetization (c and d) of the instantaneous eigenstates during the optimized dynamics in a 2D lattice.
	The population of the eigenstates in the evolving system state is indicated in color,
	showing that the population of undesired eigenstates remains negligibly small.}
	\label{inst_mag}
\end{figure}

With a finite laser amplitude, many crossings in the level spectrum of Fig.~\ref{manybodylevel} turn into avoided crossings. Changing the detuning corresponds to sweeping through these avoided crossings, resulting in increasing or decreasing occupations of the (instantaneous) eigenstates $| e_{k} \rangle$ that participate each avoided crossing. An optimized dynamics is thus characterized by a sequence of transitions ending up with population of only the desired eigenstates. Fig.~\ref{inst_mag} depicts the time-dependent energies $E_k$ (panels $a$ and $b$) and magnetization $\mathcal{M}_k(t) = \langle e_{k}|\sum_i \sigma^z_i| e_{k} \rangle$ (panels $c$ and $d$), {\it i.e.} expectation value of the collective spin operator, of the instantaneous eigenstates for the 2D lattice dynamics with optimized driving\footnote{Corresponding plots for other lattice dimensions are provided in Sec.II of ~\cite{supp}.}.
The population $|\langle e_k|\Psi(t)\rangle|^2$ of each eigenstate in the system state $\ket{\Psi(t)}$ is indicated by color with the highest population depicted in blue and lowest population depicted in yellow. The optimized dynamics is mostly supported by few instantaneous eigenstates; it thus avoids the population of undesired eigenstates to a large extent and manages to adjust the population of the desired states.

Fig.~\ref{inst_mag}(c) shows that starting from an initial state with maximally negative magnetization, the magnetization of the dominant eigenstate is increasing, while its occupation remains close to unity. Only shortly before the magnetization of the highly populated eigenstate crosses the zero-magnetization-line, a second eigenstates becomes sizeably populated. Once the occupation of these two eigenstates is approximately balanced, both states evolve to their final, {\it i.e.} maximally positive and negative magnetization. The optimal formation of the state $\ket{\Psi}$ is fundamentally different. Apart from the spurious excitation of a second eigenstate, the dynamics is supported completely by one single eigenstate that evolves continuously from the initial separable state to the final entangled state.

\begin{figure}[t!]
	\centering
	\includegraphics[width=1.0\columnwidth]{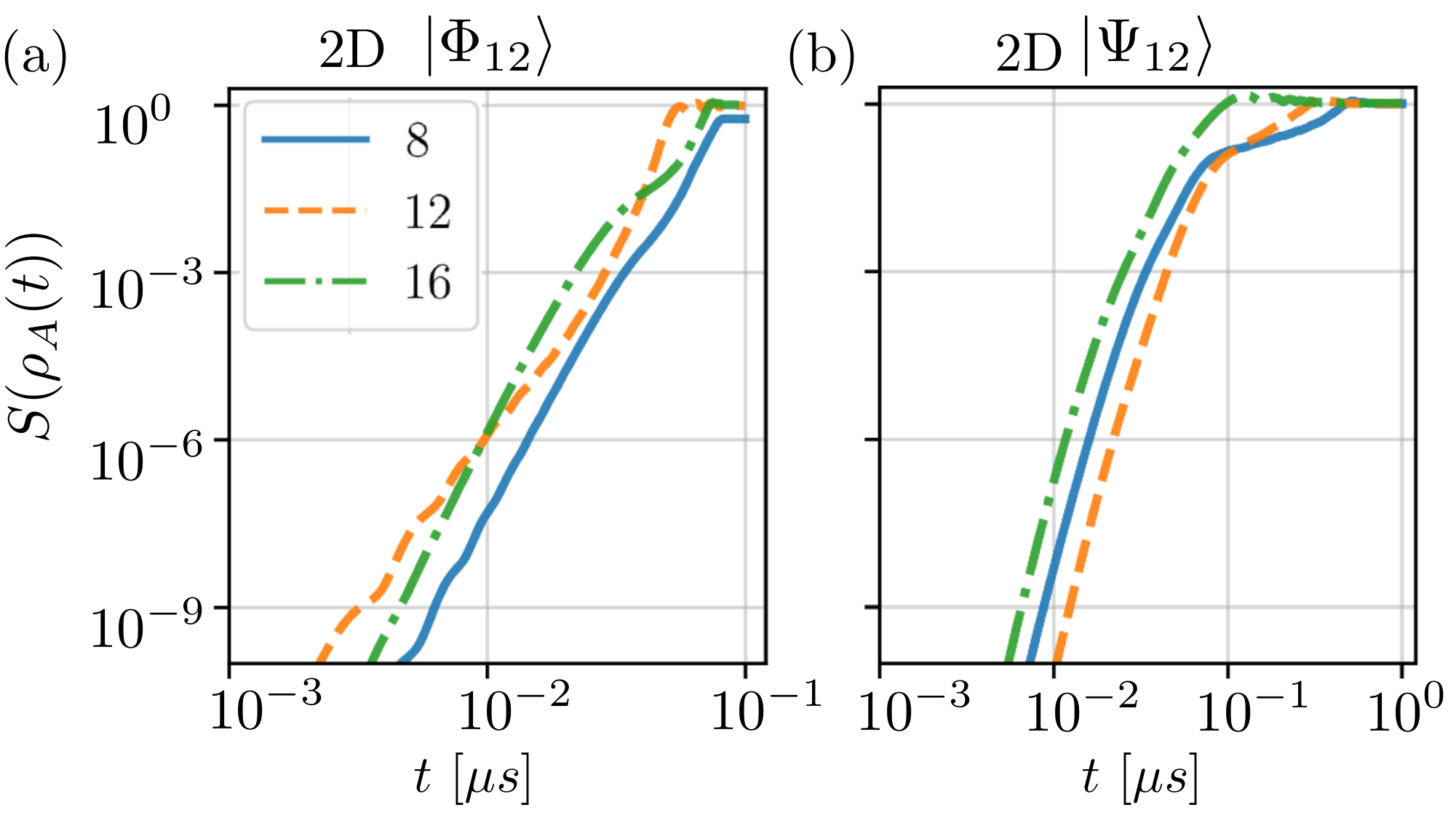}
	\caption{Dynamics of the entanglement entropy in a 2D lattice with $N=8$,$12$ and $16$ Rydberg atoms.
	The growth is ballistic with a rate that hardly depends on the system size.}
	\label{ententropy}
\end{figure}

Even though the results of Bayesian optimization (and essentially any numerical optimization procedure) depend on the choice of initialization of the search for an optimized control pulse, none of the features identified in Fig.~\ref{inst_mag} depend on this initialization, which gives strong indication that all the features are indeed essential for the optimal state preparation. This optimality is also nicely corroborated by the dynamics of the v. Neumann entropy $S(\rho_r)= -\text{Tr}[\rho_r \log_2 \rho_r]$, with the reduced density matrix $\rho_r$ of half the system. The details of the dynamics hardly depends on which spins are being traced over and the growth is essentially ballistic as shown in Fig.~\ref{ententropy}. The growth -- and thus the time-scale required for the state preparation -- is almost independent of the system size, highlighting the potential for control of large systems.

\begin{figure}[t!]
	\centering
	\includegraphics[width=1.0\columnwidth]{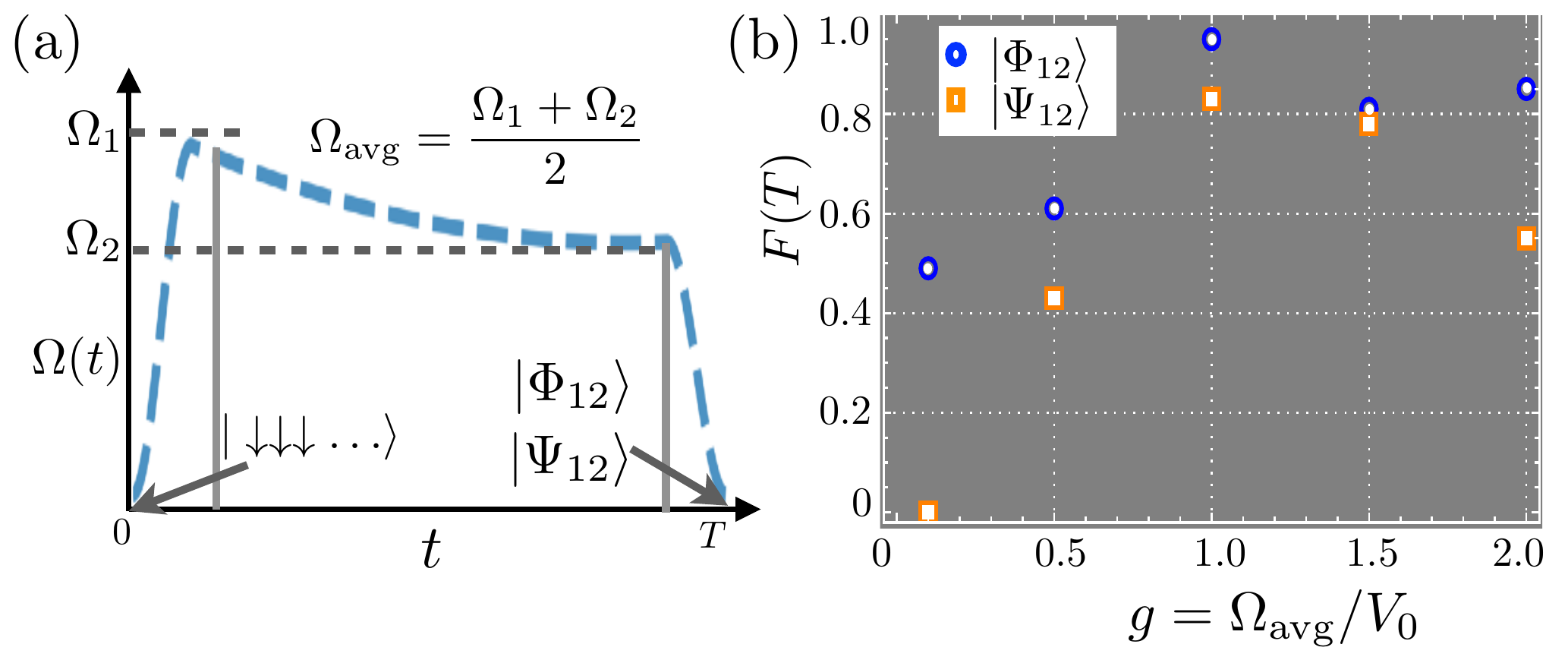}
	\caption{(a) Schematics of a typical, optimized Rabi protocol including two time-windows of quenched dynamics.
	(b) Fidelities obtained for different values of $g$, indicating that highest fidelities are obtained when $g=1$ is approached.}
	\label{quench}
\end{figure}

With these observations on magnetization and entanglement growth, one can obtain valuable intuition about the state preparation in terms of quench dynamics. As comparison of the six different Rabi profiles in Fig.~\ref{fiddyn} indicates, any such profile can be broadly separated into three stages as shown in Fig.~\ref{quench}(a): an initial quench from zero-field to $\Omega_1$, gradual change from $\Omega_1$ to $\Omega_2$ and finally a quench down from $\Omega_2$ to zero. As shown in Fig.~\ref{quench}(b), the best fidelities are obtained if $g=\Omega_{\rm avg}/V_0$ with the average Rabi-frequency $\Omega_{\rm avg} = (\Omega_1+\Omega_2)/2$ and the largest interaction constant $V_0=\max_{ij}V_{ij}$ of the Ising Hamiltonian in Eq.~\eqref{Ising}, approaches the value of $1$. This condition characterizes the quantum critical point of the transverse Ising model in 1D, and since the detunings in the protocols shown in Fig.~\ref{fiddyn} are at most comparable to the interactions, one can neglect the single atom $\sigma_z$-terms in Eq.~(\ref{Ising}) within approximation, so that the system Hamiltonian reduces to this integrable model.

The most efficient way to generate entanglement for a one dimensional transverse Ising model with short to intermediate range interactions is by quenching the dynamics close to the quantum critical point~\cite{Schachenmayer1}, and the optimized Rabi-profiles in Fig.~\ref{fiddyn}, all show this qualitative feature. The optimization has thus identified a protocol combining the ballistic dynamics of the non-integrable Ising model ~\cite{Kim} and the integrable aspect of the dynamics for values of $g$ close to 1. Strikingly, the  same functionality is also obtained in the presence of noise\footnote{Results using noisy data are presented in Sec.III  of \cite{supp}.}. Even with noise levels of several percent in Rabi-frequency and detuning, the optimizer rapidly converges to optimal solutions. The optimal driving patterns obtained deliver fidelities of around $0.95$ for higher dimensional lattices and even $0.98$ in the case of $|\Psi\rangle$. This highlights not only the importance of  performing the dynamics in higher lattice dimensions but also the fact that the state $|\Psi\rangle$ is by far, more resilient to noise than state $|\Phi\rangle$.

So far we have discussed the preparation of GHZ states in theoretical simulations, but, in particular, its noise-resilience, makes the present approach perfectly suited for working directly with experimental data~\cite{Henson2018,Nakamura2019,Duris2020}. In optimizations based on laboratory experiments\footnote{Discussion on experimental realisation of the setup is given in Sec.IV of \cite{supp}}, however, the question of how to assess the fidelity experimentally during such an optimization requires some care. The fidelity (Eq.~\eqref{fid}) is comprised not only of the directly measurable populations $\varrho_{\alpha\alpha}, \varrho_{\beta\beta}$, but also of the off-diagonal element $\varrho_{\alpha\beta}$ that can be estimated in terms of a Ramsey experiment \cite{RM2,Sommer}. The qualitative verification of coherence in terms of an interference contrast can be converted into a rigorous bound on $\varrho_{\alpha\beta}$, based on a minimal Ramsey sequence with only two data points, provided the dynamics satisfies the following conditions: in addition to coherent coupling between the states $\ket{\alpha}$ and $\ket{\beta}$, there may be dephasing in the Hamiltonian eigenbasis, population decay of the states $\ket{\alpha}$ and $\ket{\beta}$ to any other state but $\ket{\alpha}$ and $\ket{\beta}$, and any coherent and incoherent dynamics in the subspace of the system Hilbert space orthogonal to $\ket{\alpha}$ and $\ket{\beta}$. 
Any dynamics of this kind can never result in a growth of the generalized Bloch vector\footnote{Derived in Sec.V of~\cite{supp}} with elements $S_{j}(t) = {\rm Tr}[\varrho(t)\Sigma_{j}]$ defined in terms of the  Pauli-like matrices $\Sigma_x=(\ket{\alpha}\bra{\beta}+\ket{\beta}\bra{\alpha})/\sqrt{2}$ and
$\Sigma_y=i(\ket{\alpha}\bra{\beta}-\ket{\beta}\bra{\alpha})/\sqrt{2}$, as well as
$\Sigma_\alpha=\ket{\alpha}\bra{\alpha}$ and $\Sigma_\beta=\ket{\beta}\bra{\beta}$.
This results in the bound for the off-diagonal element $\varrho_{\alpha\beta}(t_i)$ in terms of population measurements performed at two different times $t_i$ and $t_f>t_i$ during  a Ramsey experiment,
\be
2|\varrho_{\alpha\beta}(t_i)|^2\ge S_\alpha^2(t_f)+S_\beta^2(t_f) - S_\alpha^2(t_i)- S_\beta^2(t_i) .
\ee
For the perfect GHZ state at instance $t_i$, one has $S_\alpha^2(t_i)=S_\beta^2(t_i)=1/4$.
A state at instance $t_f$ with $\varrho_{\alpha\beta}(t_f)=0$ and $S_\alpha(t_f)=1-S_\beta(t_f)=1$ (or $0$) then results in the bound $|\varrho_{\alpha\beta}(t_i)| \ge 1/2$. Since $1/2$ is the maximally achievable value, any reduction of observed value can thus be attributed to imperfect state preparation resulting from imperfect pulses or decoherence.

With the ability to efficiently assess the fidelity experimentally, optimal control with probabilistic machine learning techniques is likely to allow us exceeding size and quality of states that can be created, not just for GHZ states but potentially for any desired target state. In particular, an experiments with a two-dimensional lattice, identified as optimal, is likely to advance the standards in state preparation substantially with ramifications for applications including precision sensing or also measurement-based quantum computation. Apart from the explicit driving patterns identified here, the present discussion applies equally well to any other set of Rydberg states, or many other quantum systems including trapped ions, ultra-cold molecules, NV centers and superconducting qubits that can realize an Ising type Hamiltonian.

The intuitive understanding of the optimized dynamics in terms of quenches of free dynamics also holds the potential to advance our ability to develop optimal quenches for the exploration of non-equilibrium many-body phenomena in ultracold atoms or magnetic and thermodynamic behaviour of strongly correlated systems. 

We are grateful for stimulating discussions with Fr\'ed\'eric Sauvage and Robert L\"{o}w. The project Theory-Blind Quantum Control TheBlinQC has received funding from the QuantERA ERA-NET Cofund in Quantum Technologies implemented within the European Union’s Horizon 2020 Programme and from EPSRC under the grant EP/R044082/1.

\bibliographystyle{apsrev}
\bibliography{GHZref}

\clearpage

\hrulefill
\begin{center}
	\Large\textbf{Supplemental material: Bayesian optimal control of GHZ states in Rydberg lattices}
\end{center}

Figure and equation numbers in this supplemental material have an additional symbol `S'. All figures and equations referred to with a number without the additional symbol `S' are in the main manuscript.

\section{Details on Bayesian optimization of control pulses}

Bayesian optimization (BO) consists of three simple steps: (i) Build a probabilistic model that connects the figure of merit with its input parameters, (ii) Based on this model, the choice of the next set of input parameters is turned into a conditional decision problem while incorporating an incentive to explore the parameter space, (iii) Re-evaluate the figure of merit based on the new parameters and update the model. Steps (i)-(iii) form the essence of BO which are performed iteratively till convergence is achieved. In comparison to other optimal control methods, the probabilistic modeling of the optimization landscape is more suitable for noisy data and is efficient with convergence. Further technical details about BO can be found in \cite{frazier2018tutorial, brochu2010tutorial}. 

The time dependent control fields, $(\Omega(t), \Delta(t))$ are parametrized in terms of their value at $t_1=T/4$, $t_2=T/2$ and $t_3=3T/4$. For all intermediate times, the control fields are defined as quadratic splines matching these three values and the boundary conditions $\Omega(0)=\Omega(T)=0$. The search for optimal values of $\Omega(t_j)$ and $\Delta(t_j)$ (with $j=1,2,3$) is performed with Bayesian optimization in terms of the readily usable implementation~\cite{gpyopt2016}. All optimizations in the main manuscript are based on a total number of $300$ iterations, out of which $24$ are used for initialization of the optimization.
The objective of optimization (acquisition function) is the expection of improvement, and numerical integration of the Schr\"{o}dinger equation is realized with the Quimb package~\cite{Gray}.

\section{Characterizing for 1D and 3D lattice dynamics}

Fig.~\ref{supp1} depicts energy and magnetization of the instantaneous eigenstates for 1D and 3D geometries, similar to Fig.~(3) which depicts the same information for a 2D geometry. The suppression of undesired eigenstates is less effective in the 1D geometry which explains the lower fidelities obtained in this case.
\renewcommand{\thefigure}{S1}
\begin{figure}[htb!]
	\centering
	\includegraphics[width=0.9\columnwidth]{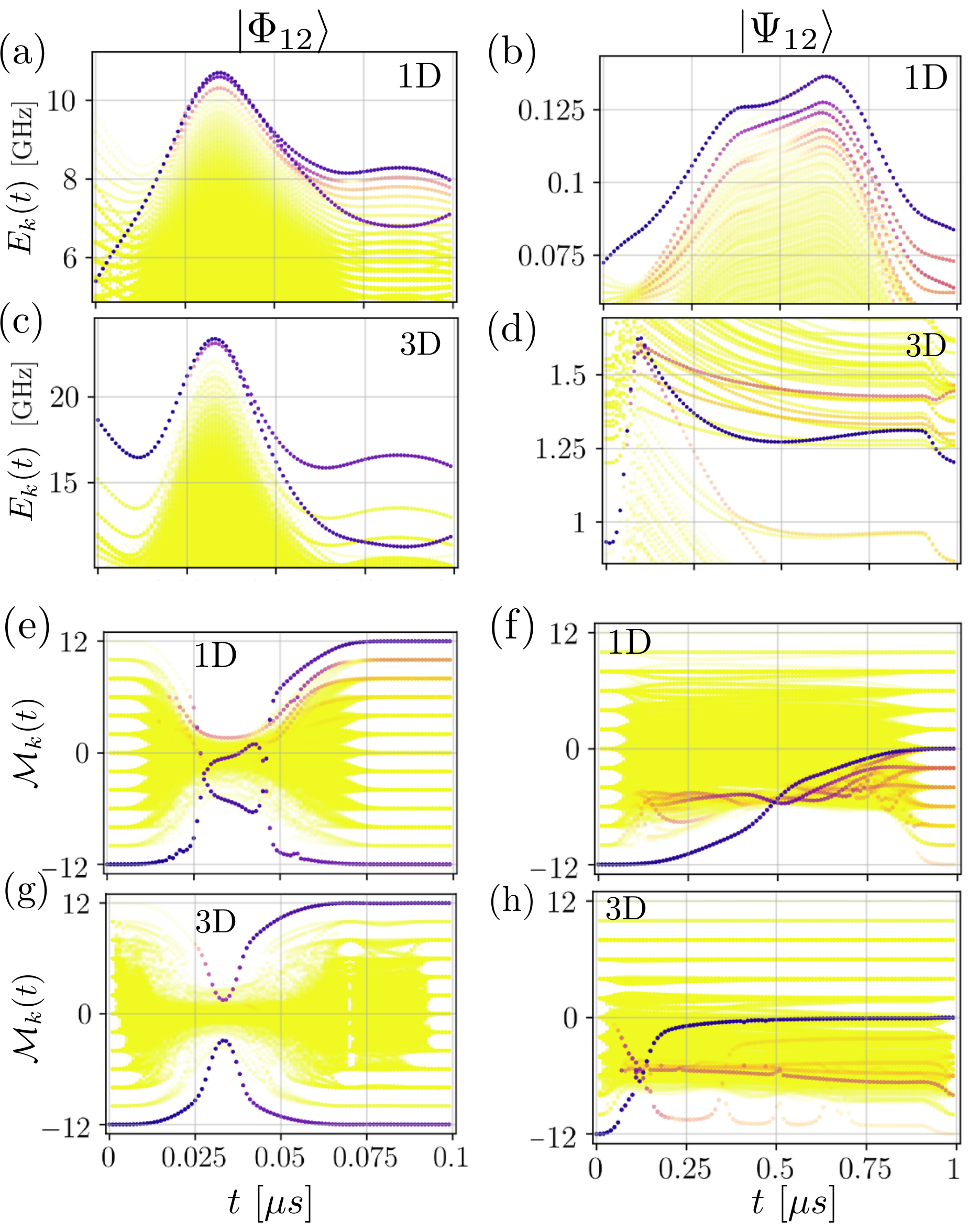}
	\caption{Energy (a to d) and magnetization (e to h) of the instantaneous eigenstates during the optimized dynamics in a 1D (a,b,e and f) and in a 3D (c,d,g and h) lattice. The population of the eigenstates in the evolving system state is indicated in color in the same way as in Fig.(3), showing that the population of undesired eigenstates remains negligibly small.}\label{supp1}
\end{figure}

Fig.~\ref{supp2} depicts the growth of entanglement entropy for $N=12$ Rydberg atoms in a 1D, 2D and 3D geometry. The ballistic growth identified in Fig.(4) for 2D systems is found also in 1D and 3D systems.
\renewcommand{\thefigure}{S2}
\begin{figure}[htb!]
	\centering
	\includegraphics[width=0.9\columnwidth]{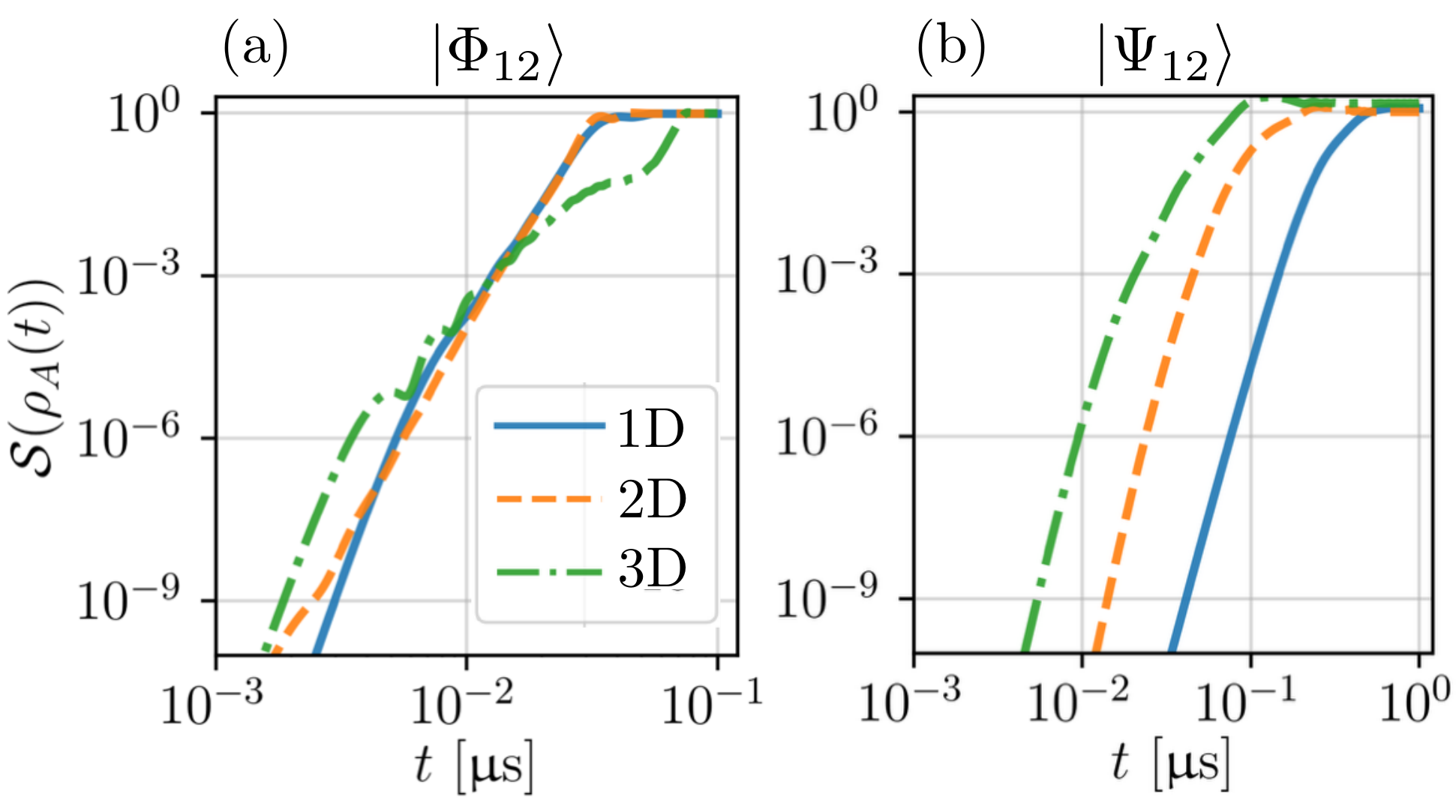}
	\caption{Dynamics of the entanglement entropy in a 1D,2D and 3D lattice with $N=12$ Rydberg atoms.}
	\label{supp2}
\end{figure}

\section{Optimizing the control pulses with noisy dynamics}

Typical experimental imperfections such as fluctuations in control fields can be taken into account in an optimization based on an ensemble-averaged quantum state. Fig.~\ref{supp3} depicts the results of such optimizations with $3\%$ fluctuations in Rabi-frequency and detuning, and the ensemble average realized in terms of $30$ ensemble members. Strikingly, the reduction in infidelity as compared to the noiseless case depends on target state and interaction geometry. While the worst is given by the target state $\ket{\Phi}$ with a drop in fidelity to about $50\%$, the reductions in 2D and 3D are substantially lower, and in the case of the target state $\ket{\Psi}$ in 2D there is essentially no reduction in fidelity.

\renewcommand{\thefigure}{S3}
\begin{figure}[htb!]
	\centering
	\includegraphics[width=0.9\columnwidth]{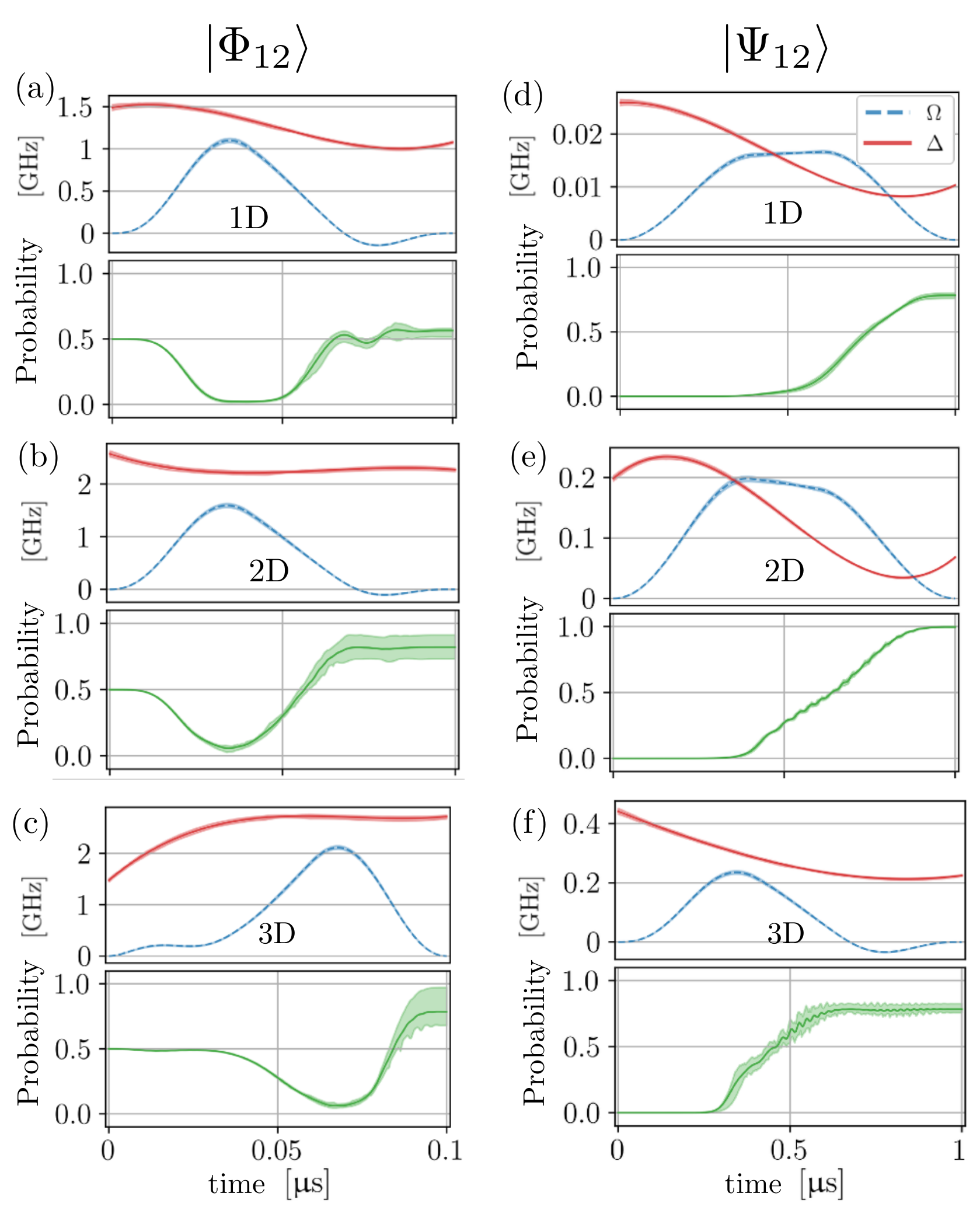}
	\caption{Results of optimized protocols with noisy laser parameters for different lattice dimensions. The mean values are depicted by solid (or dashed) lines while the shaded portions represent the corresponding variances arising from individual noisy runs.}
	\label{supp3}
\end{figure}

\section{Discussion on physical realization of the setup}

\textit{Lifetimes:} The lifetime of the Rydberg state including black-body radiation for the chosen state, 50S is about $\tau_{\rm ryd} \simeq 65$ $\si{\micro\second}$ \cite{Saffman_rev}. Since the duration $1~\si{\micro\second}$ and $0.1~\si{\micro\second}$ required for the state preparation with the present control scheme is substantially shorter, it is indeed justified to neglect any decoherence effect, arising either due to spontaneous decay or motional dynamics.

\textit{Interactions:} The ground state energy is set to zero and for the chosen Rydberg state $50S$, the van der Waals coefficient is $C_6(n) = 1.56\times 10^{-26}$ Hz m$^6$  \cite{Reinhard}, giving repulsive interactions. Our general results would just as well hold for attractive interactions. Although not included here, the Zeeman degeneracies in the interactions can introduce additional complexities but can also be incorporated.

\textit{Trapping lattices:} Atoms are assumed to be trapped in a deep optical lattices  with lattice spacing $l =1.5~\si{\micro}$m with uniform unit filling. Apart from fixing the interactions, the additional advantage of having a lattice is that it can be tuned to avoid unwanted molecular resonances. Optical lattices for ground state atoms are available in all dimensions. However, the same lattices in general do not trap Rydberg atoms resulting in unnecessary losses of atoms that would affect the overall fidelity. Nevertheless, simultaneous trapping of ground and Rydberg state atoms are conceivable using magic wavelength lattices \cite{Zhang}.

\textit{Optical parameters:}  A uniform excitation profile is assumed such that all atoms experience the same Rabi coupling at any given time. The Rabi frequencies required for the preparation of the state $\ket{\Psi_{12}}$ are an order of magnitude smaller than those for
the preparation of the state $\ket{\Phi_{12}}$, leading to larger blockade radius \cite{Saffman_rev} for the GHZ state $\ket{\Psi_{12}}$ which consists of excitations of alternate atoms compared to $\ket{\Phi_{12}}$.

\textit{Limitations on system size:} One can anticipate that the preparation of GHZ state is getting increasingly difficult with growing number of atoms $N$, because of the decreasing gaps between neighboring energy levels. Ideally, the preparation needs to be realized on time-scale shorter than $\tau_{\rm ryd}/N$, which becomes harder to satisfy with increasing $N$. For the state $\ket{\Phi_{N}}$, however, the relevant crossing shifts to larger values of detuning as $N$ is increased. This implies that the transition is separated from undesired levels. As long as sufficiently strong laser intensities are available, the preparation of the state $\ket{\Phi_{N}}$, thus seems feasible even with very large systems.

\section{Experimentally accessible estimate of state fidelity}

The bound on the off-diagonal element $\varrho_{\alpha\beta}$ of the system state given in Eq.(4) can be derived under the assumption that the dynamics is induced by a generator comprised of the following terms:
\begin{itemize}
	\item[(i)] a term describing coherent coupling between the states $\ket{\alpha}$ and $\ket{\beta}$;
	\item[(ii)] a term for dephasing in the Hamiltonian eigenbasis;
	\item[(iii)] a term for population decay of the states $\ket{\alpha}$ and $\ket{\beta}$ to any other state but $\ket{\alpha}$ and $\ket{\beta}$;
	\item[(iv)] a term for coherent and incoherent dynamics in the subspace of the system Hilbert space orthogonal to $\ket{\alpha}$ and $\ket{\beta}$. 
\end{itemize}

The four-dimensional generalized Bloch vector $\vec S$ has the following properties under the dynamics induced by the individual terms:
\begin{itemize}
	\item[(i)] the length of $\vec S$ remains invariant;
	\item[(ii)] the components $S_x$ and $S_y$ of $\vec S$ may decrease but not increase, and the components $S_\alpha$ and $S_\beta$ remain invariant;
	\item[(iii)] all components of $\vec S$ may decrease but not increase;
	\item[(iv)] all components of $\vec S$ remain invariant.
\end{itemize}

Neither of the four terms can thus induce a dynamics that results in an increase of the lengths of the generalized Bloch vector $\vec S$.
Since the Trotter decomposition guarantees that dynamics induced by the sum of the four terms is equivalent to a sequence of dynamics induced by the individual terms, the lengths of $\vec S$ can also not increase under the full dynamics including all the four terms.

For two instances in time $t_i$ and $t_f>t_i$, this thus implies
\renewcommand{\theequation}{SE1}
\be
\sum_{j=x,y,\alpha,\beta}S_j^2(t_i) \ge \sum_{j=x,y,\alpha,\beta}S_j^2(t_f)\ .
\label{eq:Sbound}
\ee
The right-hand-side of the above equation still depends on the components $S_x$ and $S_y$ which are not directly observable in practice. A practically observable bound is obtained however in terms of the worst-case assumption that $S_x(t_f)$ and $S_y(t_f)$ vanish,
\renewcommand{\theequation}{SE2}
\bqa
\sum_{j=x,y}S_j^2(t_i) &\ge& \sum_{j=x,y,\alpha,\beta}S_j^2(t_f)-\sum_{j=\alpha,\beta}S_j^2(t_i)\ ,\nonumber\\
&\ge& \sum_{j=\alpha,\beta}S_j^2(t_f)-\sum_{j=\alpha,\beta}S_j^2(t_i)\ ,
\eqa
which is equivalent to Eq.(4) with $|\varrho_{\alpha\beta}(t_i)|=\sqrt{(S_x^2(t_i)+S_y^2(t_i))/2}$.

\end{document}